\documentstyle[12pt]{article}
\newcommand{\be}{\begin{eqnarray}}
\newcommand{\ee}{\end{eqnarray}}
\newcommand{\ba}{\begin{array}}
\newcommand{\ea}{\end{array}}
\newcommand{\nn}{\nonumber}
\newcommand{\ra}{\rightarrow}

\newcommand{\LQCD}{\Lambda_{QCD}}
\newcommand{\pl}{\left|\vec{p}\,\right|}
\newcommand{\as}{\alpha_{\kappa,\tau}}
\newcommand{\ep}{\varepsilon}
\newcommand{\PP}{\mathop{\rm P}\nolimits}
\newcommand{\tf}[2]{{\textstyle\frac{#1}{#2}}}

\def\o{\over} \def\i{\infty}
\def\l{\lambda}
\def\b{\beta}
\def\bl{\b_{\l}}
\def\bm{\b_M}
\def\f{\varphi}
\def\G{\Gamma}
\def\gf{\gamma_\f}

\def\p{\partial}

\def\k{\kappa}
\def\s{\scriptstyle}
\def\ss{\scriptscriptstyle}
\def\bp{{\bf p}}
\def\kp{\kappa_{\s p}}
\def\cp{C_{\s p}}
\def\e{{\rm e}}
\def\mzz{\mid z-z'\mid}

\def\loop#1{\hbox{$\raisebox{0.8pt}{$\bigcirc$}\kern-9pt{#1}\kern+5pt$}}

\begin{document}
\begin{flushright}
ICN-UNAM-96-11\\
10th October 1996
\end{flushright}
\vskip 0.5 truein
\begin{center}
{\LARGE
Why Two Renormalization Groups are Better than One.}
${}$\footnote{This paper is based on an invited talk 
given at RG '96, Dubna, Aug. '96}
\vskip 0.4truein
{\large
\bf  C.R.\ Stephens}${}$\footnote{email: stephens@roxanne.nuclecu.unam.mx}.\\
\vskip 0.25truein
{Instituto de Ciencias Nucleares, U.N.A.M., }\\
{A. Postal 70-543, 04510 Mexico D.F., Mexico.}.\\

\end{center}
\vskip 1.3truein
{\bf Abstract:}
The advantages of using more than one renormalization group (RG)
in problems with more than one important length scale are 
discussed. It is shown that: i) using different RG's can lead 
to complementary information, i.e. what is very
difficult to calculate with an RG based on one flow parameter
may be much more accesible using another; ii) using more than
one RG requires less physical input in order to describe via RG
methods the theory as a function of its parameters; iii) using more 
than one RG allows one to solve problems with more than one 
diverging length scale.
The above points are illustrated concretely in the context of both particle
physics and statistical physics using the techniques of
environmentally friendly renormalization. Specifically, finite 
temperature $\l\f^4$ theory, an Ising-type system in a film geometry, 
an Ising-type system in a transverse magnetic field,
the QCD coupling constant at finite temperature and the crossover 
between bulk and surface critical behaviour in a semi-infinite 
system are considered.
\vfill\eject

\section{Introduction}

The renormalization group (RG), through its development in various
different guises over the last 50 years, has turned out to be one of the most 
important and powerful tools available in the study of a multitude of
different physical phenomena. The proceedings of the RG conferences
\cite{rgconf} give testament to the widespread
applicability of RG methodology. It may be felt that the RG is a well
developed and mature field with nothing fundamentally new to be learned.
I feel that such a point of view is mistaken and wish in this article
to draw attention to an area of RG usage that has been paid little attention:
that of using more than one RG. 

The principle area of interest where using more than one RG is particularly
useful is that of crossover phenomena, the hallmark of which is the 
existence of more than one asymptotic region as a function of scale wherein
the relevant effective degrees of freedom of the system are qualitatively
different.
A classic example of such behaviour is that of liquid helium confined between
two plates so as to form a film geometry \cite{gasp}. As a 
function of $L/\xi_L$,
where $L$ is the film thickness and $\xi_L$ the correlation length in the 
transverse dimensions, there are two different fluctuation dominated 
asymptotic regimes. The first is when $\xi_L/a$, where $a$ is the lattice 
spacing, and $L/\xi_L$ are both large,
the corresponding critical behaviour being that of a three dimensional system. On the
other hand when $\xi_L/a$ is large but $L/\xi_L$ is small the critical 
behaviour is that characteristic of a two dimensional system. The difference
between this and critical phenomena in an infinite space 
is that in the latter there is only really
one dimensionless ratio of interest, $\xi_L/a$, whereas in the finite 
system there
are two, the former and $L/\xi_L$. It is the existence of more 
than one scaling 
variable that makes crossovers such a rich phenomenon, both experimentally
and theoretically.

Tackling crossover problems with many conventional RG methods is 
often problematical, if
not impossible, the reason being that approximation techniques are 
more often than
not associated with an expansion around a particular fixed 
point. An intrinsic feature 
of crossover systems, however, is the existence of more than one fixed point. 
One thus requires an approximation scheme that is capable of encompassing more
than one fixed point. Such an approximation scheme has 
been developed explicitly
for a fairly large class of crossovers (though the general 
methodology should be applicable
to any crossover) under the epithet --- ``environmentally 
friendly renormalization''  \cite{NucJphysa}-\cite{runtemp} ---
which derives from the fact that many crossovers of 
interest can be fruitfully thought
of as being due to the effects of some ``environmental'' 
parameter, a prime example 
being the film thickness, $L$, above. The idea is that as the relevant 
effective degrees of freedom of a system exhibiting a crossover 
change radically as a function of scale relative to the scales set 
by the ``environment'', i.e. the effective degrees of freedom
depend on the environment, then any renormalization scheme that 
purports to describe
the crossover must perforce depend on the environment aswell. 
The aim of this paper is not to discuss at length the formalism of 
environmentally friendly renormalization but to discuss the 
advantages of using more than one RG. However, given
that more than one RG dictates that there be more than one relevant scale, and 
therefore the possibility of crossover, it is very natural to discuss multiple 
environmentally friendly RG's.

Considerations of more than one RG have been mainly confined 
to quantum field theory at finite 
temperature \cite{prevfintemp,qcd,runtemp,FTRG} and to that of an 
$O(N)$ model below the critical temperature where there exist two correlation 
lengths --- transverse and longitudinal. In \cite{gold1} the fixed point structure 
of the coupling was considered. Using two RG's, one with respect to a fiducial 
value of the transverse mass and one with respect to the longitudinal mass, 
it was possible to access the line of fixed points associated with the coexistence
curve and the critical point. Recently a more extensive analysis of the Goldstone 
problem has been carried out in four dimensions using RG's based on modifications
of minimal subtraction \cite{gold2}.

The format of the paper will be as follows: in section two I will make some 
general comments, without proof, and observations about some of the generic
advantages to be gained from using more than one RG.
In section three I consider finite temperature field theory/quantum
ferromagnets/finite size effects by considering a RG that uses as running parameter
a fiducial value of the finite temperature/size mass. In section four I will consider the 
same problem but with an RG that uses an arbitrary fiducial temperature as 
running parameter, comparing the advantages and disadvantages of the two groups.
In section five I will consider the use of a momentum and a temperature RG together in the 
context of the magetic sector of QCD. In section six the crossover between bulk and
surface critical behaviour of an Ising model in a semi-infinite geometry will be studied
as it illustrates how more than one RG may be of use in a system where there is
more than one diverging length scale. Finally in section seven I will make some conclusions
and point to some possible future work.

\section{General Considerations}

In this section I will make some general observations, without supplying
formal derivations, associated with the use of more than one RG. In particular I will
try to motivate why using more than one RG may be useful. First let us recall what it
means to use a single RG. Consider a field theory described by a 
Hamiltonian/Lagrangian (the former in the context of statistical mechanics and
the latter in the context of quantum field theory), $H=H[\{g^i_B\}]$, which is 
a function of a set of parameters $\{g^i_B\}$: momenta, sources, masses, coupling constants, 
background fields, ``environmental'' variables etc. One may think of the field
theory as being defined by these parameters and their conjugate operators.
Of interest is how the theory behaves as a function of  ``scale'', and more often than not 
in particular how the relevant effective degrees of freedom of the system change
as a function of scale, where by scale one often means as a function of one of the
parameters. 

The parameters $g_B^i$ can be related, directly or indirectly,
to the correlation functions of the theory at some particular scale, normally a
microscopic one associated with the cutoff where a mean-field analysis would be
valid. At such scales the correlation functions are very simple functions of the 
parameters. For instance, in $\l\f^4$ theory the bare coupling $\l_B$ is just the 
four point vertex function in the mean field approximation. However, well away from
this regime, where fluctuations dominate, the vertex functions will be complicated
functions of the $g_B^i$. Moreover, an approximate calculation of them, 
for example via perturbation theory in one or more of the $g_B^i$, in a
fluctuation dominated regime leads to extremely poor results (apparently infinitely
poor results in the case of relativistic quantum field theory!) 

If one thinks of the $g_B^i$ as coordinates one can ameliorate or sometimes even
circumvent completely the above problems via a coordinate change on the space 
of parameters. One defines a coordinate transformation, often by the use of a set
of normalization conditions on a relevant subset of the correlation 
functions, at a certain, arbitrary scale, $\k$. Very often the new 
parameters are simply related to the correlation functions at the scale $\k$. One 
then rewrites the actual correlation functions of the system of interest in terms of the 
new ``renormalized'' coordinates $g^i\equiv g^i(\k)$. The result is a one-parameter
family of coordinate transformations which have a group structure. Although formally
nothing can depend on this change of coordinates, i.e. the underlying physical 
theory is exactly invariant under this group of reparametrizations, when it comes to
implementing an approximation scheme, primarily perturbation theory, it is found,
as we shall see, that some coordinate systems are decidedly better than others. 

One can think of this reparametrization invariance as a global ``gauge'' invariance, 
where gauge now takes on a more literal meaning, as it was thought of originally, in
terms of ``calibration''  invariance. By going to renormalized coordinates one chooses
to calibrate the physics of a system of interest in terms of the parameters
associated with the same system but at some arbitrary, fiducial scale $\k$. The scale 
may be associated with many different quantities, some being more physical than others.
For instance, one might decide to parametrize the physics of interest of a system with
correlation length $\xi$ in terms of the parameters associated with a similar system but
with correlation length $\k^{-1}$. Another example to be treated later would be to describe
a system at temperature $T$ in terms of parameters associated with a system at a 
fiducial temperature $\tau$. Of course, $\k$ may enter in a less readily 
interpretable way such as in minimal subtraction.

Now, as mentioned, the RG is generically a one parameter group of reparametrizations
of the parameters that describe a physical system. The infinitessimal generator of this
flow is simply the vector field $\k(d/d\k)$, or in a coordinate basis $\k(\p/\p\k)+\beta^i(\p/\p g^i)$.
If there are $n$ independent parameters, one of which will be used as flow parameter, 
and therefore $n$ beta functions, some of which may of course be trivial, then in order to solve
the flow equations it is necessary to specify $n-1$ initial conditions per flow line.  
Geometrically, once an initial condition has been chosen one 
is restricted to the flow line corresponding to that initial point, 
i.e. the RG cannot be used 
to get from one flow line to another. Hence, to span the space of parameters by the RG flows
one needs an $n-1$-dimensional set of initial conditions. For instance, considering the 
four point function in $\l\f^4$ theory at the critical point as a function of two momentum variables
$p_1$ and $p_2$; then if one chooses a fiducial value of $p_2$ as flow parameter, $\k$, then
it is possible to use the RG associated with $\k$ to generate the behaviour of the four point 
function as a function of $p_2$. However, one must use as initial condition the four point 
coupling at some initial value of $\k$, $\k_0$, and at some value 
of $p_1$. The latter will be
constant along a particular flow line. This means that in order to generate $\G^{(4)}(p_1,p_2)$
one needs as physical input $\l(p_1,\k_0)$. One cannot generate $\G^{(4)}(p_1,p_2)$ purely
through use of this RG with input $\l(p'_1,\k_0)$ where $p'_1\neq p_1$.
Of course one can change initial condition but this requires
further input, i.e. knowledge of $\l(p_1,\k_0)$. As we will see in many systems such further information might not be readily available.

Another potential drawback along similar lines stems from using the RG to map to a place in
the space of parameters where a perturbative treatment is more trustworthy. 
This is often done
by choosing the RG scale such that in units of this scale the inverse correlation length is of order
one, thereby circumventing infrared difficulties. However, in crossover problems 
there are generally many length scales of interest, and in particular there may 
exist more than one relevant correlation length that may diverge. Thus fixing 
the RG scale such that one correlation length is of order unity will not necessarily 
help in the regime where the other correlation lengths are large.  

The above are potential serious complications that arise due to using an RG analysis based
on a single flow parameter. Such complications can in principle be overcome by considering an
analysis based on more than one RG. The rest of this paper will give concrete examples
of more than one RG in action in certain specific contexts. As already stated the 
intention here is to point out some general features. If one extends 
the fundamental relation between bare and renormalized vertex
functions to the case of $k$ ($k<n$) RG flows one generates $k$ RG equations for the
$N$-point vertex functions based on the reparametrization invariance of the bare theory
\be
\k_j{d\G^{(N)}\o d\k_j}(\{g^i(\{\k_j\})\},\{\k_j\})={N\o2}\gf^j(\{g^i(\{\k_j\})\},\{\k_j\})
\G^{(N)}(\{g^i(\{\k_j\})\},\{\k_j\})\label{multirg}
\ee
where $1\leq j\leq k$, $\gf^j=d\ln Z_{\f}/d\ln \k_j$, and the derivatives are taken at
fixed values of the bare parameters. One now has a set of $k$ 
vector fields which are assumed to 
be independent. As the bare theory is invariant under a Lie dragging along any of these
flows then it is also invariant under the action of the vector field $[\k_jd/d\k_j,\k_id/d\k_i]$.
This integrability condition leads to non-trivial relations between the beta functions of the
theory. In a coordinate representation each vector field can be decomposed as
\be
\k_j{d\o d\k_j}=\k_j{\p\o \p\k_j}+\beta^j_i{\p\o\p g^i}\label{multivec}
\ee
where $\b^j_i=\k_jdg^i/d\k_j$. Integrability requires that 
\be
\k_i{d\b^i_j\o d\k_i}=\k_j{d\b^j_i\o d\k_j}\label{intcond}
\ee
In the case of two RG's for example the solution of equation (\ref{multirg}) is
\be
\G^{(N)}(\{g^i(\k_1^0,\k_2^0)\},\k_1^0,\k_2^0)=\e^{{N\o2}\int_{\k_1^0}^{\k_1}
\gf^1(x_1,\k_2^0){dx_1\o x_1}}\e^{{N\o2}\int_{\k_2^0}^{\k_2}
\gf^2(\k_1,x_2){dx_2\o x_2}}\G^{(N)}(\{g^i(\k_1,\k_2)\},\k_1,\k_2)\label{multirgsol}
\ee

A key property of the RG is that if one is interested in a region, such as near a 
second order phase transition, where naive perturbation theory is 
invalid one may use reparametrization invariance to map to a 
coordinate system where perturbation theory
is better behaved by a particular choice of scale $\k$. In the case of a simple second order
phase transition where the only relevant parameter is temperature this can be achieved 
by choosing $\k$ such that in units of $\k$ the correlation length is small. A 
perturbative expansion of $\G^{(N)}(\k)$ is then valid. However, in the case when 
the correlation functions depend on many parameters it may occur that perturbation
theory is badly behaved as a function of more than one parameter. The use of a single
RG scale might be sufficient to map to a subspace of the parameter space which is
amenable to a perturbative treatment but generically there will exist other regions that
are not perturbatively accessible. By using more than one RG one will have more flexibility
in finding a region that is perturbatively accessible. For instance, considering again 
the case of two momentum variables, $p_1$ and $p_2$; assume that the correlation functions
exhibit different singular behavour in the limits $p_1\ra0$ and $p_2\ra0$. A single 
RG may be used to control the singular behaviour in one limit or the other but not both.
Having two RG's enables one to choose one scale $\k_1$ to control the singular $p_1$
behaviour and the other to control the singular $p_2$ behaviour. It is worth emphasizing
that this example is not as contrived as it sounds. In deep inelastic scattering in QCD
there are logarithms that need to be summed that are functions of the two Bjorken 
variables $x$ and $Q^2$. A single RG is capable of summing one set but not the other,
the use of two RG's gives the opportunity of controlling both sets of logs.

Another important potential advantage concerns what physical input one needs to 
specify in an RG treatment. In the case of one RG, for an $n$-dimensional parameter space it was necessary to specify $n-1$ initial conditions per flow line. With $k$ RG's 
in principle one need only specify 
$n-k$ initial conditions and therefore one has access to a $k$-dimensional subspace of the
space of parameters using only one initial condition, as by specifying only one initial condition
in this $k$-dimensional space one can flow to any other point purely through solving the
RG equations. A concrete example of this will be seen in section five.

\section{Finite Temperature Field Theory: running the mass}

In this section I will consider a RG treatment of finite temperature field theory 
using the finite temperature mass (inverse correlation length) as an arbitrary RG
scale. Attention will be restricted here to $\l\f^4$ theory 
considering the Euclidean action
\be
S[\f_{B}]=\int_0^\b dt\int
d^{d-1}x\left[{1\o2}(\nabla\f_{B})^2+{1\o2}M^2_{B}\f_{B}^2+
{\l_{B}\o4!}\f_{B}^{4}\right]\label{action}
\ee
in terms of bare quantities and the inverse temperature $\b=1/T$. The above
action also corresponds to a Hamiltonian in the same universality class as that of 
a $d$-dimensional Ising model in a transverse magnetic field, $\G$ 
\cite{EnvfRG,hertz} and of an Ising model in a film geometry with periodic 
boundary conditions \cite{NucJphysa,EnvfRG}. Any results found in the 
context of finite temperature field theory are thus easily
translatable to that of the transverse field Ising model through the identification
$M^2_B=\G-\G_c^m$, where $\G_c^m$ is the critical transverse field in the mean 
field approximation; and in the context of the Ising model in a film 
geometry with the identification $T=L^{-1}$. 

Near a second order phase transition the finite temperature mass, $M(T)\ra0$,
which as is well known leads to severe problems in implementing perturbation theory
in the infrared. A ``standard''
renormalization which emphasizes the ultraviolet, such as minimal subtraction,
fails to remedy the situation. As the effective degrees of freedom change
qualitatively as a function of temperature, the ``environment'' here, it is wise to
use an environmentally friendly renormalization that respects this fact 
\cite{NucJphysa,prevfintemp}.
To specify an environmentally friendly set of coordinates I will use the following
normalization conditions 
\be
\left.{\p\o\p
p^2}\G^{(2)}(p,M(T)=\k,\l(\k),T,\k)\right|_{p=0}
&=&1,\\
\G^{(2)}(p=0,M(T)=\k,\l(\k),T,\k)&=&\k^2,\label{massCnd}\\
\G^{(4)}(p=0,M(T)=\k,\l(\k),T,\k)&=&\l(\k).\label{normcnds}
\ee
Here the renormalization is at a fiducial value, $\k$, of the finite temperature mass.
I will concentrate, for the purposes of illustration, on the four 
point function, which
with the above normalization conditions gives the beta function at one loop 
\be
\beta(h)=-\ep(T/\k)h+h^2+O(h^3)\label{beta}
\ee
in terms of the 
floating coupling \cite{NucJphysa,EnvfRG} defined 
by $h=-{3\o2}\l(\k)\k {d\o d\k}\loop2$, i.e. by normalizing the 
coefficient of $h^2$ in $\beta(h)$ to unity. 
The symbol $\loop{k}$ stands for the one-loop diagram with $k$ propagators,
without vertex factors, at zero external momentum.
It can be obtained from the following basic diagram in $d$ dimensions
\be
\bigcirc&=&
T\sum_{n=-\infty}^\infty\int{d^{d-1}k\o(2\pi)^{d-1}}
\ln(k^2+(2\pi n\tau)^2+\k^2)\nn\\
&=&-{\G(-{d\o2})\k^d\o{(4\pi)}^{d/2}}-{2T^d\o{(4\pi)}^{(d-1)/2}\G({d+1\o2})}
\int_0^\infty dq{q^d\o\sqrt{q^2+{z^2}}}{1\o e^{\sqrt{q^2+{z^2}}}-1},
\ee
where $z=\k/T$, by differentiations with respect to $\k^2$.
The first derivative gives ${d\bigcirc/ d\k^2}=\loop1$,
whereas for $k\geq1$ we have the general rule that the derivative
with respect to $\k^2$ of the loop with $k$ propagators
gives $-k$ times the loop with $k+1$ propagators.

The solution of (\ref{beta}) is 
\be
h(z)={e^{-\int_{\ss{z_0}}^{\ss {z}}\ep(x){dx\o x}}\o{h_0^{-1}(z_0)
-\int_{\ss{z_0}}^{\ss {z}}
e^{-\int^x_1\ep( x'){dx'\o x'}}{dx\o x}}}\label{floating}
\ee
where
\be
\ep(z)=5-d-(7-d){{\displaystyle\sum_{n=-\i}^{\i}{4\pi^2n^2\over z^2}
\left(1+{4\pi^2n^2\over z^2}\right)^{d-9\over2}}
\over{\displaystyle\sum_{n=-\i}^{\i}\left(1+{4\pi^2n^2\over
z^2}\right)^{d-7\over2}}}\label{eps}
\ee
and $z_0=T/\k_0$.
For $d<4$ one can take the limit $\k_0\ra\i$ whereupon the dependence on the
initial condition drops out completely and one is left with the universal one 
loop floating coupling
\be
h(z)=(5-d){{\displaystyle\sum_{n=-\infty}^{\infty}}{(1+{({2\pi n\o
z})}^2)}^{(d-7)\o2}
\o{\displaystyle\sum_{n=-\infty}^{\infty}}{(1+{({2\pi n\o
z})}^2)}^{(d-5)\o2}}\label{floatuni}
\ee

There are several important points to be noted here: firstly $h(z)$ 
displays three fixed points not just two as would be the case for a
beta function based on a non-environmentally friendly renormalization
scheme such as minimal subtraction. Besides the trivial Gaussian 
fixed point, in the limit $T/\k\ra0$, one finds a $d$-dimensional fixed 
point $h=(4-d)$ equivalent to that found using minimal subtraction. However, in the
limit $T/\k\ra\i$, which physically corresponds to approaching a second 
order phase transition, one finds a $d-1$-dimensional fixed point $h=(5-d)$
which shows that near the phase transition there is a dimensional reduction.
In terms of the dimensionful coupling $\l$ the above analysis shows that
$\l\ra M(T)/T$ in four dimensions as $M(T)\ra0$. Of course, this 
does not mean the theory becomes
non-interacting; $\loop2\ra(T/M(T))$ as $M(T)\ra0$ and hence diverges, however 
the product goes to a constant which is associated with the three dimensional 
fixed point. The floating coupling takes this into account in a very natural way.  

Thus one sees that environmentally friendly renormalization is capable of 
describing the entire dimensional crossover between $d$ and $d-1$ dimensions.
One finds that the dimensionally reduced regime can be described in terms 
of a set of critical exponents characteristic of $d-1$ dimensions. These critical 
exponents have been calculated up to two loop order of a Pad\'e-resummed 
perturbation for the coupling constant \cite{EnvfRG} and are in good agreement 
with experimental values. One also finds
that the crossover is completely universal when $T\ll\k_0$ and $M(T)\ll\k_0$.
That is, in terms of the above results, the crossover in the coupling does not
depend on $h(z_0)$. 

Given that it is possible to describe the complete crossover here 
what, if any, are the drawbacks? In the context at hand we must think in 
terms of what parameters are being used to describe the system. 
As far as the RG is concerned here $T$ is a fixed parameter and the flow is with
respect to the finite temperature mass, i.e. $T$ is a constant along any flow line. 
In the context of a quantum ferromagnet,
where $M(T)$ is a function of $\G$, one has access to another parameter, the 
transverse field, with which $M(T)$ may be changed at fixed $T$. However, 
for a relativistic field theory, such as the Higgs sector of the standard model, 
such a parameter, physically at least, does not exist. Thus in this case the flow consists of 
how the theory varies as a function of bare Higgs mass at fixed temperature.
So although one can access the same phase diagram for the quantum ferromagnet
and the Higgs model, in the former case varying the finite temperature mass is physically
quite intuitive whereas this is not so for the latter where it is much more natural to think
of relating the physics at one temperature to another temperature, zero for example. 
The other input parameter is $h(T,\k_0)$. If one is interested in only universal 
quantities, as in near
the second order phase transition above, then there will be no dependence on
$h(T,\k_0)$, however, there are also non-universal features of interest, such as the 
critical temperature, which will depend on $h(T,\k_0)$.  Thus in order to access such 
quantities at different temperatures one has to supply more information through the 
couplings $h(T,\k_0)$ at different temperatures. 

The above shows that using
an RG based on the finite temperature mass as running parameter has advantages
and disadvantages depending on the physics associated with the 
particular problem of interest. It is clear though that there exist quantities of
interest, such as the critical temperature, that cannot be accessed purely by the above
RG. All the above reasons are sufficient motivation for seeking an alternative RG that
may overcome some, if not all the above disadvantages.

\section{Finite Temperature Field Theory: running the temperature}

Given the drawbacks, at least in the context of finite temperature field theory,
of running the finite temperature mass one may ask whether one can use
another  RG to overcome this difficulty. Given that it is of interest to describe 
what happens at temperature $T$ in terms of zero temperature parameters
I will now consider using a fiducial temperature, $\tau$, as the running parameter.
Given that $\tau$ is totally arbitrary and at our discretion one generates an RG based on
the independence of the bare theory from $\tau$. Once again given that
the aim is to describe more than one asymptotic regime wherein the effective 
degrees of freedom are very different one must find an environmentally 
friendly coordinate system. This will be specified by the following normalization conditions
(I will restrict attention to the disordered phase $T>T_c$, the extension to $T<T_c$ 
can be found in \cite{runtemp})
\be
\left.{\p\o\p
p^2}\G^{(2)}(p,M(\tau),\l(\tau),T=\tau)\right|_{p=0}
&=&1,\\
\G^{(2)}(p=0,M(\tau),\l(\tau),T=\tau)&=&M^2(\tau),\label{massCndtwo}\\
\G^{(4)}(p=0,M(\tau),\l(\tau),T=\tau)&=&\l(\tau).\label{normcndstwo}
\ee
The flow functions to one loop are
\be
\bm & = &{\l\o 2} \tau{\p\loop1\o\p\tau} \\
\bl & = &-{3\o2}\l^2\tau{d\o d\tau}\loop2 .\label{couplingflow}
\ee
Note that in $\b(\l)$ it is a total derivative $d/d\tau=\p/\p\tau+\bm\p/\p M$
that appears rather than a partial derivative. This is essential because as
the critical temperature is approached both
$\l\hbox{ and }M\sim\tau-T_c$ (see below) which implies that
\be
{\bl-\bl^p\o\bl^p}={4\pi\tau\o9M}-{2\o3\pi}+\dots,\label{diffTerm}
\ee
diverges at the critical temperature, where $\bl^p$ is the beta function found
from using $\p/\p\tau$. The additional term (\ref{diffTerm}) is therefore the dominant contribution
and cannot be neglected. Thus treating $\tau dM^2/d\tau$ as being of higher order
will not be consistent.
If one had dropped this term, one would then find that the coupling
rather than going to zero approaches a constant. 
However as is known from previous work \cite{NucJphysa,EnvfRG,prevfintemp},
and as was seen in the last section,
the coupling $\lambda\rightarrow0$ as one approaches the critical point
from either above or below. In the present approach this is a direct consequence of the total
derivative in the flow function $\bl$. Of course, as also mentioned in the previous section 
the vanishing of $\l$ does not imply the theory becomes non-interacting.
In terms of the floating coupling using $M$ as a parameter 
measuring the distance from $T_c$ we obtain for $h$ as $M\ra0$ 
the flow equation 
\be
M{\p h\o\p M}=-(5-d)h+h^2+O\left(\textstyle{M\o T}\right).
\label{uFlow}
\ee
This equation has, in the limit $M\ra0$, a familiar fixed point structure,
with a stable, non-trivial fixed point at $h^*=(5-d)$.
Moreover, it has the added advantage of providing a coupling that remains
``small'' for all temperatures, and is proportional to the zero-temperature
coupling in the zero temperature limit.

The differential equation for the coupling
(\ref{couplingflow}) is easy to solve,
since it contains a total derivative, and to the order we are
working takes the same form in both phases. The solution is
\be
\l^{-1}(\tau)=\l^{-1}(\tau_0)
+{3\o2}\left[\loop2(M(\tau),\tau)-\loop2(M(\tau_0),\tau_0)\right],
\label{lambdaSolution}\ee

After solving the flow equations one is free to choose the reference
temperature $\tau$ equal to the actual temperature $T$ of interest.
In fact this is essential if one wishes to obtain perturbatively sensible
results for physical quantities.
The renormalization conditions (\ref{massCndtwo}), (\ref{normcndstwo})
show that the parameters $M(T)$ and $\lambda(T)$ describe the behaviour
of the vertex functions $\G^{(2)}$ and $\G^{(4)}$ at zero momentum.
With these equations one is also able to determine the critical temperature
in terms of the zero-temperature parameters $M(0)$ and $\l(0)$.
As may be expected on dimensional grounds $T_c$ is proportional to $M(0)$,
the constant of proportionality being a function of $\l(0)$.
If one takes values for $\l(0)$ and $M(0)$ to be those associated with
estimates for the equivalent parameters in the Higgs sector of the
standard model one finds, for $\l(0)=1.98$ and $M(0)=200$GeV
that $T_c=613$GeV. 

Since $M(T)\ll T$ near $T_c$ the finite-temperature four-dimensional theory
reduces there to a three-dimensional Landau-Ginzburg model. In 
the neighbourhood of the critical temperature the general vertex functions 
have the form
\be
\G^{(n)}_{\pm}=\gamma^{(n)}_{\pm}|T-T_c|^{\nu\left(d_c-n{d_c-2+\eta\o2}\right)},
\label{scaling}
\ee
where $d_c=d-1$ is the reduced dimension at the 
critical point, $\nu$ and $\eta$ are 
critical exponents and $\gamma^{(n)}_{\pm}$ are amplitudes.
The appearance of the critical exponent $\nu$ is unusual in particle physics.
It reflects the need for composite operator renormalization 
and the physical dependence of $M(T)$ on temperature. This ensures that
the exponent $\nu$ is physically accessible in finite temperature 
field theory whereas it is usually not experimentally observable in a 
particle physics context as the dependence of the renormalized 
mass on the bare mass is experimentally inaccessible.
As we see here, however, the exponent $\nu$ in fact plays a highly 
significant role near the critical point.

Near the critical temperature one finds 
$M^2_{\pm}={(C^{\pm})}^{-1}|T-T_c|^{\gamma}$ with
$\l_{\pm}=l^{\pm}|T-T_c|^{\nu}$ and $m_{\pm}={(f_1^{\pm})}^{-1}|T-T_c|^{\nu}$,
where the notation of Liu and Fisher \cite{LiuFisher} for the amplitudes is being used.
The temperature flow equations give values for the exponents 
$\nu=1$ and $\eta=0$ at one loop, which are not as good as those obtained 
by flowing the mass parameter. In terms of amplitude ratios one finds
\be
{C^+\o C^-}=4,\qquad\qquad
{f_1^+\o f_1^-}=2\sqrt{12\o13}\approx1.92,\qquad\qquad
{l^+\o l^-}={1\o2}.
\ee
The fact that these are not one is indicative of a 
cusp in the graphs of mass and coupling versus temperature as the theory 
passes through the critical temperature.
These ratios are universal numbers analogous to the critical exponents.
The best estimates for the amplitude ratios are the high- and low-temperature
series expansion results of Liu and Fisher \cite{LiuFisher} who find
\be
{C^+\o C^-}=4.95\pm0.15,\qquad\qquad
{f_1^+\o f_1^-}=1.96\pm0.01,\qquad\qquad
\ee
which are in good agreement with the above RG results. By comparison: 
at tree level  (mean field theory) $C^+/C^-=2$ and $f_1^+/f_1^-=1.41$, 
whilst in the $\varepsilon$ expansion at order $\varepsilon^2$, 
assuming dimensional reduction, 
$C^+/C^-=4.8$ and $f_1^+/f_1^-=1.91$.

One can see that the values found for the amplitude ratios are 
substantially better than the corresponding
results for critical exponents. This indicates a complimentarity
between the current approach of flowing the environment, temperature, and 
that of the previous section where the flow parameter was the finite 
temperature mass.  At one loop the latter group gives better results for 
exponents whereas the former gives better results 
for amplitudes, but both schemes should converge to the same results 
as one goes to higher orders. 
The crucial point here is that by using a second different RG it has been 
possible to access information that would have been very difficult to obtain
by running the RG of section 3. Here with this group one has access to 
non-universal quantities such as the critical temperature and critical 
amplitudes. It is clear then that there are distinct 
advantages in being able to implement 
more than one RG. In the next section we will consider a situation where 
two groups are used but this time at the same time.

\section{Finite Temperature Field Theory: running temperature and momentum}

In this section we will consider running two RG's in the context of
investigating the QCD coupling constant in the magnetic sector as
a function of momentum and temperature (see \cite{qcd} for more details).
We use as a renormalization condition that the static
(i.e. zero energy), spatial three-gluon vertex equals
the tree-level vertex in the symmetric momentum configuration
\be
\left.\Gamma^{abc}_{ijk}(p_i^0=0,\vec{p}_i,g_{\k,\tau},T=\tau)
\right|_{{\rm symm.}\atop\k}=
g_{\k,\tau}f^{abc}\left[g_{ij}(p_1-p_2)_k+\mbox{cycl.}\right].
\ee
In contradistiction to the previously discussed cases this chosen 
renormalization condition depends now on two arbitrary parameters,
the momentum scale $\k$, and the temperature scale $\tau$.
Therefore we can perform an RG analysis with respect
to both parameters, i.e. we can run more than one environmental
parameter at the same time.

In order to get rid of ambiguities arising from
gauge dependence the Landau gauge Background Field
Feynman rules resulting from the Vilkovisky-de Witt effective
action are used.
Due to the corresponding Ward Identities the calculation is simplified
in that one only has to calculate the transverse gluon self energy
function $\Pi^{\rm Tr}$ in the static limit.
In terms of the coupling $\as:=g^2_{\k,\tau}/4\pi^2$ the $\beta$
functions are then
\be
\k\frac{d\as}{d\k}=
\as\,\left.\pl\frac{d\Pi^{\rm Tr}}{d\pl}\right|_{\pl=\k\atop T=\tau},
\label{betaKappa}\qquad\qquad
\tau\frac{d\as}{d\tau}=
\as\,\left.T\frac{d\Pi^{\rm Tr}}{dT}\right|_{\pl=\k\atop T=\tau}.
\label{betas}
\ee
The $\tau$ RG is needed to draw conclusions
about the temperature dependence of the coupling.
This can not be done using the $\k$-scheme alone without assuming something
about the temperature dependence of the initial value of the coupling
used in solving the differential equation as was seen in section 3 in the context of $\l\f^4$.

The resulting two beta functions are
\be
\k\frac{d\as}{d\k}=\beta_{vac}+\beta_{th},\qquad\qquad
\tau\frac{d\as}{d\tau}=-\beta_{th}, \label{resultBeta}
\ee
where the vacuum contribution is, as usual,
\be
\beta_{vac}&=&\as^2\left(-\tf{11}{6}N_c+\tf{1}{3}N_f\right),
\ee
and where, in terms of the IR and UV convergent integrals
\be
F_n^\eta=\int_0^\infty\!dx\frac{x^n}{e^{\k x/2\tau}-\eta}
\left[\log\left|\frac{x+1}{x-1}\right|-
2\!\sum_{k=0}^{\tf{n}{2}-1}\frac{x^{2k+1}}{2k+1}\right]
\ee
and
\be
G_n^\eta=\int_0^\infty\!dx\,\frac{1}{e^{\k x/2\tau}-\eta}
\PP\frac{x}{(x^2-1)^n},
\ee
the thermal contribution is given by
\be
\beta_{th}&=&\as^2\left[
\left(\tf{21}{16}F^1_0+\tf{3}{4}F^1_2-
\tf{3}{2}G^1_0-\tf{25}{8}G^1_1-\tf{}{}G^1_2\right)N_c\right.+\nn\\
&&\left.\left(\tf{1}{4}F^{-1}_0+\tf{3}{4}F^{-1}_2-
\tf{3}{2}G^{-1}_0-G^{-1}_1\right)N_f\right].
\ee
Because the two beta functions (\ref{resultBeta}) are not exactly
each other's opposite the RG improved coupling
is not just a function of the ratio $\k/\tau$.
There is another dimensionful scale (such as $\LQCD$) that comes
from an initial condition for these differential equations.
The solution of the set of coupled differential equations can be written
in the form
\be
\as=\frac{1}{\left(\tf{11}{6}N_c-\tf{1}{3}N_f\right)
\ln\tf{\k}{\LQCD}-f\!\left(\tf{\k}{\tau}\right)}
\ee
where the function $f$ satisfies $\beta_{th}=\as^2\k df/d\kp$
with the initial condition $\lim_{\tau\downarrow0}f=0$ so that
one can identify $\LQCD$ with the usual zero-temperature QCD scale.
Actually this function $f$ can be found in terms of the functions $F$ and $G$:
\be
f=\left(\tf{21}{16}F^1_0+\tf{1}{4}F^1_2+\tf{7}{8}G^1_1\right)N_c+
\left(\tf{1}{4}F^{-1}_0+\tf{1}{4}F^{-1}_2\right)N_f.
\ee

The high-temperature behaviour (i.e. for $\tau\gg\k$) is determined by
\be
f\longrightarrow N_c\tf{21\pi^2}{16}\tf{\tau}{\k}+
\left(\tf{11}{6}N_c-\tf{1}{3}N_f\right)\ln\tf{\k}{\tau}+O(1).
\ee
The sign of this coefficient is such that
for increasing temperatures at fixed momentum scale one enters a 
strong coupling regime and
the coupling grows without bound. The opposite sign would lead to
asymptotic freedom in this limit as originally suggested in
\cite{Collins}. The limit $\tau/\k\ra\infty$
is an IR limit where confinement takes place, so
unless at higher loop order the magnetic mass increases quickly enough
with temperature in order to act as an effective IR cutoff, one
apparently cannot circumvent
this problem without actually solving confinement.
This is an important consideration when considering phase transitions
which involve non-abelian gauge fields.

In the regime $\tau\gg\k$ the beta functions behave as in a three-dimensional
theory so this is the region where dimensional reduction occurs.
Here it is natural, as for $\lambda\phi^4$, to use a different
dimensionless coupling $u=\as\frac{\tau}{\k}$ since then fixed points
may turn up more clearly.
However in this case such a reparametrization cannot avoid the strong
coupling region.

If one allows the momentum-scale to change with temperature simultaneously,
the high-temperature limit can be taken in many ways.
In the region $\tau\gg\k$ the shape of the constant coupling 
contours is given by $\tau\sim\k\ln\tf{\k}{\LQCD}$.
This characterizes exactly along which paths in the $(\tau,\k)$-plane
the coupling increases or decreases.
For example at a fixed ratio $\tau/\k$ (no matter what this ratio is)
one eventually finds a coupling that decreases like $1/\ln\k$,
much in the same way as at zero temperature.
This is a natural contour to consider for a weak-coupling 
regime where one could treat the quark-gluon plasma as
a perfect gas,
as then the thermal average of the momentum of massless quanta at
temperature $T$ is proportional to the temperature.
However at low momenta the assumption of weak coupling breaks down.
Furthermore, instead of considering quantities at the average momentum
it is more appropriate to use thermal averages of the quantities
themselves as a weighted integral over all momenta.
But once again one runs into problems at the low-momentum end as long
as we cannot treat the strong-coupling regime.

In this section then we see the advantages of using two independent RG's
simultaneously. Using only one initial condition, i.e. one point in the two
dimensional $p$, $T$ parameter space, it is possible to reach any other 
point purely through RG flows. 

\section{Bulk/Surface Crossover}

In this section I illustrate the use of more than one RG in a case where
there exists more than one diverging length scale --- the crossover between
surface and bulk critical behaviour in a semi-infinite system. The basics of this
subject have been covered by Hans Diehl \cite{diehl} in another article in this volume so
here I will give only a cursory treatment. Consider a $d$-dimensional ($d>2$)
semi-infinite Ising-like system described by the continuum Hamiltonian
\be
H=\int_0^{\i} dz\int
d^{d-1}x\left[{1\o2}(\nabla\f_{B})^2+{1\o2}(T-T_c^{bm})\f_B^2+{1\o2}C_B\delta(z)\f_B^2
+{\l_{B}\o4!}\f_{B}^{4}-H_B\f_B\right]\label{ham}
\ee
where $C_B$ is the bare surface enhancement parameter, $T_c^{bm}$ is the bulk mean
field critical temperature and $H_B$ is a position
dependent magnetic field that we can restrict to having only surface support if
required. The free field propagator in a mixed representation where we Fourier transform in the
infinite transverse directions is 
\be
G(z,z',\bp)={1\o2\kp}\left(\e^{-\kp\mzz}+\cp\e^{-\kp(z+z')}\right)\label{prop}
\ee
where $\kp=(\bp^2+T-T_c^{bm})^{1\o2}$ and $\cp=(\kp-\cp)/(\kp+\cp)$. The important 
observation here is that there are two types of pole associated with the propagator 
for $\bp=0$: one where $T=T_c^{bm}$ and one where $(T-T_c^{bm})^{1\o2}+C_B=0$. The
former corresponds to the bulk critical point while the second, for $C_B<0$ 
corresponds to a surface phase transition at a mean field surface critical temperature
$T_c^{sm}=T_c^{bm}+{\mid C_B\mid}^2$. Note that in this case where the surface
interactions are enhanced relative to those of the bulk the surface critical temperature is 
higher than that of the bulk and therefore the surface will order at a 
temperature $T_c^{sm}$ in the presence of a disordered bulk. 

As there are two possible, independent diverging length scales in this problem --- 
the bulk correlation length and the surface correlation length --- 
it is not clear that a RG
based on one parameter will be sufficient to give a reliable perturbative treatment of 
both phase transitions. That is to say that if one uses an RG to map 
out from the bulk critical
region to a mean field type region where a perturbative treatment should be adequate it is
not clear that this will be sufficient to cope with the badly behaved surface fluctuations. 
It is also clear that due to the inhomogeneity fluctuations are position dependent and 
therefore to implement an environmentally friendly renormalization one should have a 
set of normalization conditions that are capable of capturing the relevant position 
dependence. Given the existence of two diverging length scales and position dependence
the best course of action would seem to be to implement two environmentally friendly
RG's. To show how this works I will consider here only the case of the four point 
vertex function at one loop. Near the surface phase transition one ought to find a $d-1$ 
dimensional fixed point for the coupling and near the bulk transition a $d$-dimensional one.
For $c<0$, where $c$ is the renormalized surface parameter, 
the normalization conditions I will use are the following:
\be
\G^{(4)}(z,\bp_i=0,\k_1,\k_2)\equiv\int dz_1dz_2dz_3
\G^{(4)}(z,z_1,z_2,z_3,\bp_i=0,\k_1,\k_2)
 = \l(z,\k_1,\k_2)\label{norm1}\\
{\p\o\p z}G^{(2)}(z,z'=0,\bp=0,\k_1,\k_2) = 
-\k_1G^{(2)}(z,z'=0,\bp=0,\k_1,\k_2)\hskip 0.3truein\label{norm2}\\
\G^{(2)}(z=\i,z'=\i,\bp=0,\k_1=0,\k_2) = 0\hskip 1truein\label{norm3}\\
\G^{(2)}(z=0,z'=0,\bp=0,\k_1,\k_2=0) = 0\hskip 1truein\label{norm4}
\ee
The conditions (\ref{norm3}) and (\ref{norm4}) serve to fix the shifts in the bulk critical
temperature and the surface enhancement respectively. 
The second condition determines the wavefunction renormalization to be
\be
Z_{\f}(z,\k_1,\k_2)=1+{\l\o8}\int{d^{d-1}p\o(2\pi)^{d-1}}{\cp\o\kp^2}{\e^{-{2\kp z}}\o(\kp+\k_1)}\label{wfnorm}
\ee
Note that it is explicitly $z$ dependent \cite{ocstup}.
As there are two RG scales one can naturally introduce two dimensionless couplings
$\l_1=\l\k_1^{d-4}$ and $\l_2=\l\k_2^{d-4}$ with beta functions
\be
\k_i{d\l_i\o d\k_i}=-(4-d)\l_i+2{\gf}^i\l_i-3\l_i^2\k_i^{4-d}
F_i(z,\k_1,\k_2)\label{betafn}
\ee
where
\be 
F_i=\int_0^{\i}dz' {d^{d-1}p\o(2\pi)^{d-1}}
G^{(2)}(z,z',\bp=0,\k_1,\k_2)\k_i{dG^{(2)}\o d\k_i}(z,z',\bp=0,\k_1,\k_2)\label{betafnn}
\ee
and ${\gf}^i=d\ln Z_{\f}/d\ln\k_i$. 

As I am considering a situation where the surface orders before the bulk I will be 
interested mainly in how the four point function varies as a function of the surface 
correlation length. In the limit $z\ra\i$, $\beta_2\ra -(4-d)\l_2$ 
which just shows that $\l$ scales with its canonical dimension with 
respect to $\k_2$ as one would expect. In the limit $z\ra0$, $\k_2\ra0$ 
for $\k_1$ fixed, i.e. as the surface transition is approached one finds
\be
\k_2{d\l_2\o d\k_2}=-(4-d)\l_2+3\l_2^2S_{d-1}\G({d-1\o2})\G({7-d\o2}){\k_1\o\k_2}\label{asbeta}
\ee
where $S_{d-1}$ is the volume of the $d-1$-dimensional sphere.
Bearing in mind that the surface exhibits $d-1$-dimensional behaviour the natural 
$d-1$-dimensional dimensionless coupling is $u_2=\l_2(\k_1/\k_2)$ which satisfies
\be
\k_2{d u_2\o d\k_2}=-(5-d)u_2+3u_2^2S_{d-1}\G({d-1\o2})\G({7-d\o2}){\k_1\o\k_2}\label{asbetau}
\ee
which yields precisely the $d-1$-dimensional Wilson-Fisher fixed point. 
In terms of the floating coupling one finds
\be
\k_2{dh\o d\k_2}=-(4-d^2_{\ss\rm eff})h+h^2\label{float}
\ee
where the effective dimension $d^2_{\ss\rm eff}$ is 
\be
d^2_{\ss\rm eff}=d+\k_2{d\o d\k_2}\ln F_2(z,\k_1,\k_2)\label{deff}
\ee
which in the limit $(\k_2/\k_1)\ra0, \ \k_1z\ra0$ yields $d^2_{\ss\rm eff}\ra d-1$.
The anomalous dimension $\gf^2$ is 
\be
\gf^2=\l(\k_1,\k_2){\k_2^{6-d}\o 2(\k_1^2-\k_2^2)^{1\o2}}S_{d-1}
\int_0^{\i}dp{p^{d-2}\e^{-\kp z}\o \kp(\kp+\k_1)(\kp+(\k_1^2-\k_2^2)^{1\o2})^2}\label{gefftwo}
\ee
Near the surface phase transition, $z\ra0$, $\k_2\ra0$ one finds 
$\gf^2\ra \l_2(\k_1,\k_2)(\k_2/\k_1)$. As 
$\l_2(\k_1,\k_2)(\k_2/\k_1)\sim (\k_2/\k_1)$ one
sees that the anomalous dimension of the field at the surface goes to zero 
as we know it must at the one loop level in a $d-1$-dimensional $\l\f^4$ theory. 

To describe the extraordinary
transition one can use the above techniques but with the added complication that the bulk is
ordering in the presence of an already ordered surface and therefore one must use a propagator
associated with a $z$-dependent magnetization. Conceptually this is not difficult but makes 
for much more complicated calculations. For $c>0$ there is only one diverging length scale, the
bulk correlation length, therefore a single RG will suffice. One 
finds that the normal $d$-dimensional Wilson-Fisher fixed point is the 
only non-trivial one. More details will be given in a future publication.

\section{Conclusions}

In this paper I have tried to give a flavour for why one should consider using more than 
one RG, especially in considering crossover systems. The principal advantages are:
i) access to complementary information, as shown in the example of comparing finite
temperature field theory with a running finite temperature mass RG and with a 
running temperature RG; ii) less physical input and therefore more predictive power. 
This was illustrated in the case of finite temperature QCD where with two RG's specification
of the coupling at one momentum and one temperature was sufficient for the RG to be able to
calculate  the coupling at any other momentum or temperature. In the case of one RG
it would be neccessary to have a line of initial conditions; iii) more flexibility in finding a 
perturbatively treatable region of parameter space in problems with more than one diverging
length scale as was seen in section six.

Without doubt there are very many problems where the use of more than one RG would 
facilitate a solution, here I will mention just a few of relevance. Certainly the case of 
the crossover between bulk and surface critical behaviour deserves a much more detailed
analysis to be able to access full crossover scaling functions for both the surface and
extraordinary transitions. A more detailed analysis of the $O(N)$ theory would also be 
useful. In terms of particle physics the standard model could be examined as a function of
the Higgs mass and the mass of the vector bosons for instance; also deep inelastic scattering could be examined as a function of $Q^2$ and $x$, the use of two 
RG's being sufficient to resum
the logs associated with both variables. Finally, there are many interesting questions to be
answered about the methodology of using more than one RG such as: are the beta 
functions always integrable in perturbation theory? are there any other consistency 
requirements for using more than one RG etc.?

\section*{Acknowledgements}

This work was supported by Conacyt under grant number 211085-5-0118PE.
I wish to thank Denjoe O'Connor and Chris Ford for conversations; Axel Weber
for a reading of the manuscript and the JINR, Dubna for hospitality.

\end{document}